\documentclass{aastex}
\usepackage{spr-astr-addons}
\usepackage{url}
\RequirePackage{color}

\newcommand{\al}{\alpha}

\newcommand{\om}{\Omega}

\newcommand{\be}{\begin{equation}}
\newcommand{\ee}{\end{equation}}
\newcommand{\n}{\label}

\newcommand{\ben}{\begin{eqnarray}}
\newcommand{\een}{\end{eqnarray}}
\begin{document}
\title{Crossing the phantom divide with a classical Dirac field}
\slugcomment{Not to appear in Nonlearned J., 45.}
%% Running heads
\shorttitle{Short article title}
\shortauthors{Autors et al.}
\author{Mauricio Cataldo\altaffilmark{1}}
\affil{Departamento de F\'\i sica, Facultad de Ciencias, Universidad
del B\'\i o--B\'\i o, Avenida Collao 1202, Casilla 5-C,
Concepci\'on, Chile.\\} \and
\author{Luis P. Chimento\altaffilmark{2}}
\affil{Departamento de F\'\i sica, Facultad de Ciencias Exactas y
Naturales, Universidad de Buenos Aires, Ciudad Universitaria
Pabell\'on I, 1428 Buenos Aires, Argentina.}

\altaffiltext{1}{mcataldo@ubiobio.cl}
\altaffiltext{2}{chimento@df.uba.ar}

\begin{abstract}
In this paper we consider a spatially flat
Friedmann-Robertson-Walker (FRW) cosmological model with
cosmological constant, containing a stiff fluid and a classical
Dirac field. The proposed cosmological scenario describes the
evolution of effective dark matter and dark energy components
reproducing, with the help of that effective multifluid
configuration, the quintessential behavior. We find the value of the
scale factor where the effective dark energy component crosses the
phantom divide. The model we introduce, which can be considered as a
modified $\Lambda$CDM one, is characterized by a set of  parameters
which may be constrained by the astrophysical observations available
up to date.
\end{abstract}


%\section*{}
%\label{sec:intro}

\section{Introduction}
According to the standard cosmology the total energy density of the
Universe is dominated today by both dark matter and dark energy
densities. The dark matter, which includes all components with
nearly vanishing pressure, has an attractive gravitational effect
like usual pressureless matter and neither absorbs nor emits
radiation. The dark energy component in general is considered as a
kind of vacuum energy with negative pressure and is homogeneously
distributed and, unlike dark matter, is not concentrated in the
galactic halos nor in the clusters of galaxies. The observational
data provide compelling evidence for the existence of dark energy
which dominates the present--day Universe and accelerates its
expansion.

In principle, any matter content which violates the strong energy
condition and possesses a positive energy density and negative
pressure, may cause the dark energy effect of repulsive gravitation.
So the main problem of modern cosmology is to identify the form of
dark energy that dominates the Universe today.

In the literature the most popular candidates are cosmological
constant $\Lambda$, quintessence and phantom matter. Their equation
of state is given by $w=p/ \rho$, where $w=-1$, $w>-1$ and $w<-1$
respectively. Dark energy composed of just a cosmological term
$\Lambda$ is fully consistent with existing observational data.
However, these data do not exclude the possibility of explaining the
observed acceleration with the help of phantom matter. The
cosmological constant can be associated with  a time independent
dark energy density; the energy density of quintessence scales down
with the cosmic expansion, and the energy density of phantom matter
increases with the expansion of the Universe.

Mostly, the attention has been paid to dark energy as high energy
scalar fields, characterized by a time varying equation of state,
for which the potential of the scalar field plays an important role.
Among scalar field models we can enumerate quintessence
models~\citep{Uno1,Uno2,Uno3}, Chameleon fields~\citep{Dos1,Dos2},
K-essence~\citep{Myrzakulov1,Tres1,Tres2,Tres3,Tres4,Tres5},
G-essence~\citep{Yerzhanov,Kulnazarov}, Chaplygin
gases~\citep{Cuatro1,Cuatro2,Cuatro3},
tachyons~\citep{Cinco1,Cinco2,Cinco3}, phantom dark
energy~\citep{PhDE}, cosmon models~\citep{Bauer,Grande1,Grande2},
etc.

%(Despues agrego las referencias indicadas mas abajo)
In general the crossing of the phantom divide cannot be achieved
with a unique scalar field~\citep{seis1,seis2,seis3}. This fact has
motivated a lot of activity oriented towards different ways to
realize such a
crossing~\citep{11-29A,11-29B,11-29C,11-29D,11-29F,11-29GG,11-29G,11-29H,11-29I,11-29IA,11-29J,11-29K,11-29L,11-29M,11-29N,11-29O,11-29Q,11-29R,11-29S,11-29T,11-29U,11-29V,
11-29W,11-29X,11-29Y,11-29Z,11-29ZZ,11-29ZZZ,11-29ZZZZ,11-29GGH,11-29GGHH}.
For instance, in Ref.~\citep{Ruth} was explored the so called
kinetic k-essence
models~\citep{32-36A,Cinco3,32-36C,32-36D,32-36E,Tres4}, i.e.
cosmological models with several k-fields in which the Lagrangian
does not depend on the fields themselves but only on their
derivatives. It was shown that the dark energy equation of state
transits from a conventional to a phantom type matter. Note that
formally, one can get the phantom matter with the help of a scalar
field by switching the sign of kinetic energy of the standard scalar
field Lagrangian~\citep{PhDE}. So that the energy density
$\rho_{ph}=-(1/2)\Phi^2+V(\Phi)$ and the pressure
$p_{ph}=-(1/2)\Phi^2-V(\Phi)$ of the phantom field leads to
$\rho_{ph}+p_{ph}=-\Phi^2<0$, violating the weak energy condition.

In the Universe nearly $70\%$ of the energy is in the form of dark
energy. Baryonic matter amounts to only $3-4\%$, while the rest of
the matter (27 $\%$) is believed to be in the form of a non-luminous
component of non-baryonic nature with a dust--like equation of state
($w = 0$) known as cold dark matter (CDM). In this case, if the dark
energy is composed just by a cosmological constant, then this
scenario is called $\Lambda$CDM model.

%\newpage

Below, we analyze a FRW universe having cosmological constant and
filled with a stiff fluid and a classical Dirac field (CDF). With
this matter configuration, we will see that the FRW universe evolves
from a non--accelerated stage at early times to an accelerated
scenario at late times recovering the standard $\Lambda$CDM
cosmology. The CDF may be justified by an important property: in a
spatially flat homogeneous and isotropic FRW spacetime it behaves as
a ``perfect fluid" with an energy density, not necessarily positive
definite. This pressureless ``perfect fluid" can be seen as a kind
of ``dust". In particular, motivated by the fact that the dark
matter is generally modelled as a system of collisionless
particles~\citep{DMA,DMB,DMC,Copeland}, we have the possibility of
giving to cold dark matter content an origin based on the nature of
the CDF. On the other hand, the stiff fluid is an important
component because, at early times, it could describe the shear
dominated phase of a possible initial anisotropic scenario,
dominating the remaining components of the model.

The organization of the paper is as follows: In Sec. II we present
the dynamical field equations for a FRW cosmological model with a
matter source composed of a stiff fluid and a CDF. In Sec. III the
behavior of the dark energy component is studied. In Sec. IV we
conclude with some remarks.

%%%%%%%%%%%%%%%%%%%%%%%%%%%%%%%%%%%%%%%%%%%%
\section{Dynamical field equations}
%%%%%%%%%%%%%%%%%%%%%%%%%%%%%%%%%%%%%%%%%%%%%

We shall adopt a spatially flat, homogeneous and isotropic spacetime
described by the FRW metric
\par
\be \label{Metrica} ds^{2} = dt^{2} -
a^{2}(t)\left(dx^{2}+dy^{2}+dz^{2}\right),
\end{equation}
\noindent where $a(t)$ is the scale factor. The spacetime contains a
cosmic fluid composed by (i) a stiff fluid $\rho_s=\rho_{s0}/a^6$
and (ii) a homogeneous classical Dirac field $\psi$. The
Einstein-Dirac equations are \ben \label{00}
3H^2-\Lambda=\frac{\rho_{s0}}{a^6}+\rho_{D}, \\
%\label{m} \dot\rho_M+3H\rho_M=0, \\
\label{DF} \left(\Gamma ^{i}\nabla _{i} - \al\right)\psi  = 0, \een
\noindent where $H=\dot{a}/a$  is  the  Hubble  expansion rate,
$\alpha$ is a constant  and the dot denote differentiation with
respect to the cosmological time. Here $\rho_{D}$ represents the
energy density of the CDF.

The  dynamical equation for the CDF in curved spacetime can be
obtained using the vierbein formalism. So, $\Gamma ^{i}$  are the
generalized  Dirac  matrices, which satisfy the anticommutation
relations \be \{\Gamma^i,\Gamma^k\}=-2g^{ik}I, \ee with the metric
tensor $g^{ik}$ and $I$ the identity $4\times 4$ matrix. They can be
defined in terms of the usual representation of the flat space-time
constant Dirac matrices $\gamma^i$ as \be \n{dm} \Gamma^0=\gamma^0,
\qquad \Gamma^\beta=\frac{\gamma^\beta}{a}, \ee where the Dirac
matrices $\gamma^i$ can be written with the Pauli matrices
$\sigma^\beta$ as
\begin{eqnarray}
\gamma^0=i \left(\begin{array}{cc}
I&0\\
0&-I\\
\end{array}\right),
\qquad \gamma^\beta=i \left(\begin{array}{cc}
0&-\sigma^\beta\\
\sigma^\beta&0\\
\end{array}\right)
\n{gai}
\end{eqnarray}
and
\begin{eqnarray}
\sigma_1= \left(\begin{array}{cc}
0&1\\
1&0\\
\end{array}\right),
%\quad
\sigma_2= \left(\begin{array}{cc}
0&-i\\
i&0\\
\end{array}\right), \nonumber \\
%\quad
\sigma_3= \left(\begin{array}{cc}
1&0\\
0&-1\\
\end{array}\right),
\n{pm}
\end{eqnarray}
with $I$ the identity $2\times 2$ matrix. The symbol $\nabla
_{i}=\partial_i+\Sigma_i$ denotes the spinorial covariant
derivatives, being the spinorial connection $\Sigma_i$ defined by
$\nabla_i\Gamma_k=0$. Then it leads to \be \n{sc} \Sigma_0=0, \qquad
\Sigma_\beta=\frac{1}{2}H\Gamma^0\Gamma_\beta. \ee

Coming back to the Dirac equation (\ref{DF}), it takes the form \be
\n{de}
\left[\Gamma^0\left(\partial_t+\frac{3}{2}H\right)-\al\right]\psi(t)=0.
\ee Restricting ourselves  to  the metric (\ref{Metrica}), the
general solution of the latter equation consistent with Eq.
(\ref{00}) is given by
\par
\ben \psi (t) =\frac{1}{a^{3/ 2}} \left(\begin{array}{c}
   b_{1}\,e^{-{\rm i}{\rm \al}t}\\
   b_{2}\,e^{-{\rm i}{\rm \al}t}\\
   d^{*}_{1}\,e^{\,{\rm i}{\rm \al}t}\\
   d^{*}_{2}\,e^{\,{\rm i}{\rm \al}t}\\
\end{array}\right)
\n{spinor} \een \noindent with arbitrary complex coefficients
$b_{1}$, $b_{2}$, $d_{1}$ and $d_{2}$. The only nonvanishing
component of the energy-momentum tensor for the CDF is
\par
\begin{equation}\label{TD}
T^{{{\rm D}}}_{00} = {{\rm \al}\over a^{3}}\left(|b_{1}|^{2} +
|b_{2}|^{2} - |d_{1}|^{2} - |d_{2}|^{2}\right) \equiv
{\rho_{{D0}}\over a^{3}},
\end{equation}
where $\rho_{D0}=\al(b^2-d^2)$, $b^2=|b_{1}|^{2} + |b_{2}|^{2}$ and
$d^2=|d_{1}|^{2} + |d_{2}|^{2}$. For positive values of
$\rho_{{D0}}$ this source formally behaves as a perfect fluid
representing a classical  dust. However, the CDF will allow us to
extend the analysis for negative values  of the energy density.
Here, we restrict ourselves to the physical sector $d^2<b^2$, this
means that $\rho_D \geq 0$.

From Eq.~(\ref{TD}) we see that the energy density of the CDF is
given by $\rho_D=\rho_{D0}/a^3$. Thus we shall rewrite the Friedmann
equation~(\ref{00}) in the following form:
\begin{eqnarray}\label{FRD}
3H^2=\rho_{m} +\rho_x,
\end{eqnarray}
where
\begin{eqnarray}
\n{1}
\rho_{m}=\frac{ \al b^2}{a^3}, \\
\n{2} \rho_x=\frac{\rho_{s0}}{ a^6}-\frac{\al d^2}{a^3}+\Lambda.
\label{rho} \een Here $\alpha$, $b^2$ and $d^2$ have dimensions of
time$^{-1}$.

Notice that there is some arbitrariness in the distribution of the
four components of Eq. (\ref{TD}) between dark matter and dark
energy, since it is not clear that we can represent each dark
component with a mix of positive and negative energy components. To
avoid this arbitrariness we have included both positive energy
components of CDF into dark matter and both negative energy
components of CDF into dark energy. This assignment has an invariant
meaning because it is preserved under linear transformations of CDF.

This distribution allows us to interpret the positive part of the
Dirac energy--momentum tensor as a pressureless matter (giving rise
to the total (``true") observable matter $\rho_m$); while its
negative part, together with the stiff fluid and the cosmological
constant, we interpret as the effective dark energy component
$\rho_x$. It must be noted that this is equivalent to assume that we
have a dust component $\rho_M =\rho_{M0} \, a^{-3}$ and a CDF
$\rho_{D}=\rho_{D0} \, a^{-3}$ associated with a negative energy
density (i.e. with $d^2>b^2$) in the dark energy sector.

The general solution of the Einstein equation (\ref{FRD}) with
sources (\ref{1}) and (\ref{2}) takes the form
\begin{eqnarray}
a^3(t)=\frac{\alpha (b^2-d^2)}{2 \Lambda}\left[-1 +\cosh
\sqrt{3\Lambda}t\right] \nonumber \\
+\sqrt{\frac{\rho_{s0}}{\Lambda}}\sinh \sqrt{3\Lambda}t,
\end{eqnarray}
where we have set the initial singularity at $t=0$. In
Fig~\ref{scalef} is shown the behavior of the scale factor and its
derivatives.

\begin{figure}
\plotone{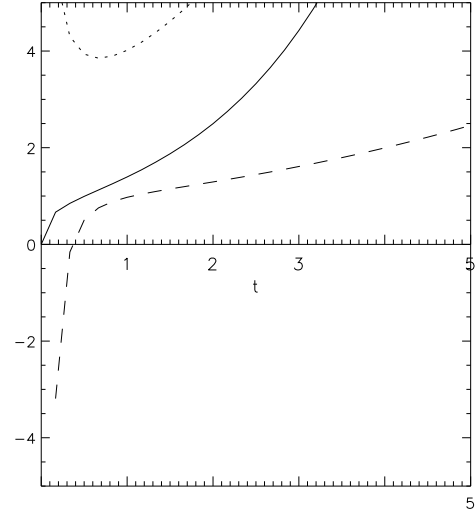}
\caption{\label{scalef} We show the behavior of the scale factor
(solid line) and its derivatives $\dot{a}$ (dotted line) and
$\ddot{a}$ (dashed line). From this it is clear that this
cosmological scenario exhibits an accelerated expansion since there
is a stage where $\ddot{a}>0$.}
\end{figure}

%%%%%%%%%%%%%%%%%%%%%%%%%%%%%%%%%%%%%%%%%%%%%%%%%%
\section{Dark energy evolution}
%%%%%%%%%%%%%%%%%%%%%%%%%%%%%%%%%%%%%%%%%%%%%%%%%%
It is well known that the present observational data do not exclude
the existence of the phantom dark energy; thus it is interesting to
consider the possibility of crossing the phantom divide. We shall
see that the Dirac field component can produce a strong variation of
the dark energy density $\rho_x$, allowing one to have such a
crossing at early times.

In order to have a positive $\rho_x$, we choose the parameters of
the models according to the following restriction:

\begin{eqnarray}\label{constraint}
\al^2d^4< 4 \Lambda \rho_{s0}.
\end{eqnarray}
Since $a(t)$ is an increasing function of time, this effective dark
energy component decreases until it reaches a minimum value
$\rho_{xc}=\Lambda-\al^2d^4/4\rho_{s0}$ at $a_c=(2\rho_{s0}/\al
d^2)^{1/3}$ where the dark component crosses the phantom divide and
begins to increase with time (see Fig.~\ref{fig15}). Fundamentally,
the effective dark energy component crosses the phantom divide due
to the presence of the $d$--parameter. In fact, the cosmological
evolution of dark energy depends on the negative term $-\alpha
d^2/a^3$, see Eq.~(\ref{rho}), because it produces a minimum at
$\dot\rho_x(a_c)=0$, showing the importance of considering the CDF
as a source of the Einstein equation.

\begin{figure}
\plotone{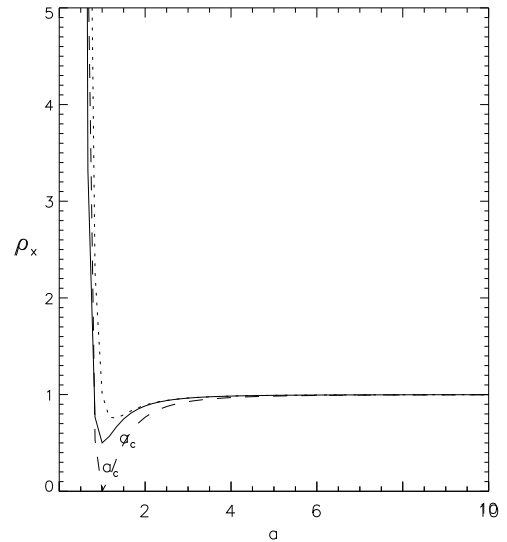}% Here is how to import EPS art
\caption{\label{fig15} We show the behavior of $\rho_x$ as a
function of time. The dashed line represents the limit case
$a_c=a_m$, i.e. $\al^2d^4= 4 \Lambda \rho_{s0}$, while solid and
dotted lines represent a typical case satisfying the condition
$a_c>a_m$ since $\al^2d^4< 4 \Lambda \rho_{s0}$.}
\end{figure}
\begin{figure}
\plotone{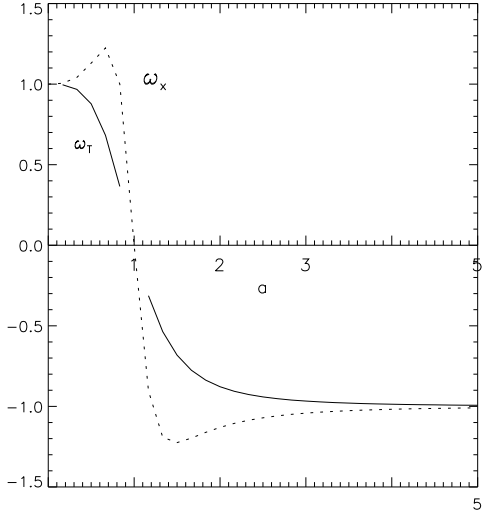}% Here is how to import EPS art
\caption{\label{stateP} We show the behavior of the dark energy
state parameter $w_x$ (dashed line) and the state parameter for the
full source content $w_{_T}$ (solid line). It is clear that the dark
energy violates the dominant energy condition while the full source
content does not violate it. Note also that the dark energy
component remains in the phantom region as it enters into it after
crossing the phantom divide.}
\end{figure}
\begin{figure}
\plotone{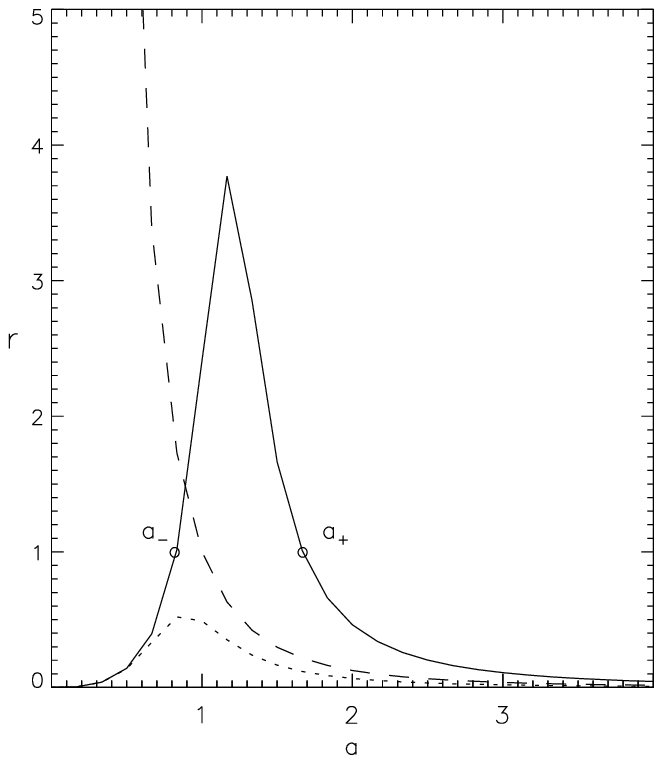}% Here is how to import EPS art
\caption{\label{fig15ratio} We show the behavior of the ratio
$r=\rho_m/\rho_x$ as a function of the scale factor $a$. The dashed
line represents the $\Lambda$CDM cosmology, where $r \propto 1/a^3$.
The solid line represents the case where $r>1$ and so at the
beginning dark energy dominates, then at the range $a \in (a_-,a_+)$
the dark matter component dominates over dark energy and, for
$a>a_+$, the dark energy component dominates again. The dotted line
represents a scenario where the dark energy always dominates over
the dark matter, since $r_{max}<1$.}
\end{figure}

Assuming that total matter and dark energy are coupled only
gravitationally, then they are conserved separately, so we have that
\ben
\dot{\rho}_m+3H \rho_m=0,\\
\label{ECDE} \dot{\rho}_x+3H (1+w_x)\rho_x=0, \een where we have
assumed the equation of state $p_x=w_x \rho_x$ for the dark energy
component. Taking into account that the dark energy component has a
variable state parameter $w_x=w_x(a)$, we define as
$a_m=(\rho_{s0}/\Lambda)^{1/6}$ the value of the scale factor where
this equation of state coincides with the matter one, that is
$w_x=0$. So that, the above restriction $\al^2d^4< 4 \Lambda
\rho_{s0}$ now becomes $a_m<a_c$. In terms of the above parameters,
$a_c$ and $a_m$, the dark energy state parameter $w_x$ can be
written as \be \label{omegax}
w_x=\frac{1-(a/a_m)^6}{1-2(a/a_c)^3+(a/a_m)^6}. \ee On the other
hand we can consider the state equation for the full source content.
For this 2--component system we define the total pressure
$p_{_T}=w_{_T} \rho_{_T}$, where $p_{_T}=p_x$ and the total energy
density $\rho_{_T}=\rho_m+\rho_x$. This implies that \be
w_{_T}=\frac{w_x}{1+r}, \ee where $r=\rho_m/\rho_x$. In
Fig.~\ref{stateP} we compare the behaviors of the state parameters
of the dark energy and the full source content.

Let us consider in more detail the ratio of energy densities of
matter and dark energy. The defined above ratio takes the form \be
\label{r} r =\frac{\alpha b^2a^3}{\Lambda a^6-\alpha d^2
a^3+\rho_{s0}}. \ee It is easy to show that this ratio has a maximum
at $a=a_c=(\rho_{s0}/\Lambda)^{1/6}$ (at this point the
$d^2r/da^2<0$). This maximum value is given by \be r_{max}=
\frac{\alpha b^2 \sqrt{\rho_{s0}/\Lambda}}{2\rho_{s0}-\alpha  d^2
\sqrt{\rho_{s0}/\Lambda}}. \ee It is clear that for cosmological
scenarios where $r_{max}<1$ the dark energy component always
dominates over the dark matter during all cosmological evolution.
Thus, in order to have stages where dark matter dominates over dark
energy, we have to require that $r_{max}>1$. So in this case we
would have two values for the scale factor where the energy density
of dark matter equals the energy density of the dark energy:
\begin{eqnarray}
a_{\pm}=\left[\frac{\alpha
(b^2+d^2)\pm\sqrt{\alpha^2(b^2+d^2)^2-4\Lambda
\rho_{s0}}}{2\Lambda}\right]^{1/3}.
\end{eqnarray}
Thus, for cosmological scenarios where $r_{max}>1$, at the beginning
the dark energy dominates over the dark matter until the scale
factor reaches the value $a=a_-$ where the dark matter energy
density equals the energy density of dark energy and then it begins
to dominate. This stage of domination of the dark matter is
prolonged until the moment when the scale factor reaches the value
$a=a_+$ and the dark energy again starts to dominate over the dark
matter (see Fig.~\ref{fig15ratio}).

%%%%%%%%%%%%%%%%%%%%%%%%%%%%%%%%%%%%%%%%%%%%%%%%%%%%%%
\subsection{Constraints on cosmological parameters}
%%%%%%%%%%%%%%%%%%%%%%%%%%%%%%%%%%%%%%%%%%%%%%%%%%%%%%

In order to confront our model and the cosmological observations we
shall use the constraints on cosmological parameters obtained from
the analysis of observational data assuming the $\Lambda$-CDM model.
This choice is justified by the fact that our model differs from
$\Lambda$-CDM only at early times. It must be noted that in general
such a procedure of using constraints derived from a fit of one
specific model can give at best rough estimations for the parameters
of a different cosmological model.

The proposed scenario is characterized by four parameters which may
be constrained by the astrophysical observations available up to
date. Since we have considered flat FRW cosmological scenarios, the
dimensionless density parameters are constrained today by
\begin{equation}\label{omega igual 1}
\Omega_{m,0}+\Omega_{x,0}=1.
\end{equation}
From Eqs.~(\ref{FRD})--(\ref{rho}) we have that
\begin{equation}\label{frdeq}
3H^2=\frac{ \al b^2 }{a^3}+\frac{\rho_{s0}}{ a^6}-\frac{\al
d^2}{a^3}+\Lambda.
\end{equation}
Evaluating it today (where we set $a=1$) we have that
\begin{equation}
\rho_{crit}=3H_0^2=\al b^2+\rho_{s0}-\al d^2+\Lambda,
\end{equation}
so the two dimensionless density parameters are given by
\begin{equation}
\n{om} \Omega_{m,0}=\frac{\rho_m (a=1)}{\rho_{crit}}=\frac{\alpha
b^2}{\al b^2+\rho_{s0}-\al d^2+\Lambda},
\end{equation}
\begin{equation}
\n{ox} \Omega_{x,0}=\frac{\rho_x
(a=1)}{\rho_{crit}}=\frac{\rho_{s0}-\al d^2+\Lambda}{\al
b^2+\rho_{s0}-\al d^2+\Lambda}.
\end{equation}
Now it may be shown that, in general, for a flat FRW cosmology the
deceleration parameter $q$ is given by
\begin{equation}
q=-\frac{\ddot{a}a}{\dot{a}^2}=\frac{1}{2}+\frac{3}{2}
\frac{p_{_T}}{\rho_{_T}}.
\end{equation}
Taking into account that we have $p_{_T}=p_x=\omega_x \rho_x$ and
$\rho_{_T}=\rho_m+\rho_x$ the deceleration parameter is given by
\begin{equation}
q=\frac{1}{2}+\frac{3}{2} \omega_{x} (1-\Omega_m),
\end{equation}
and evaluating it today (i.e. $a=1$) and using~(\ref{omegax})  we
obtain
\begin{equation}\label{qq}
\al(b^2-d^2)(1-2q_0)+2\rho_{s0}(2-q_0)=2\Lambda(1+q_0).
\end{equation}
Thus, for accelerated scenarios, $q<0$, we require a positive
cosmological constant. Another constraint may be introduced by
taking into account the moment when the Universe has started to
accelerate again. In other words, this is related to the moment when
the Universe starts violating the strong energy condition ($\rho+p
\geq 0$ and $\rho+3p \geq 0$). So we must require the inequality
$\rho+3p<0$. Now from the condition $\ddot{a}=0$ and the equivalent
Friedmann equation
\begin{eqnarray}\label{tresp}
\frac{\ddot{a}}{a}=-\frac{1}{6} (\rho+3p),
\end{eqnarray}
we conclude that $\rho+3p=0$, which implies that $\rho_{_m}+\rho_x+3
\omega_x \rho_x=0$, obtaining the condition
\begin{eqnarray}\label{zacc}
\al(b^2-d^2)(1+z_{acc})^3+4\rho_{s0}(1+z_{acc})^6=2\Lambda,
\end{eqnarray}
where the equation $1/a=(1+z)$ was used. Here $z_{acc}$ is the value
of the redshift when the Universe starts to accelerate again.

In conclusion we have the four conditions~(\ref{om}), (\ref{ox}),
(\ref{qq}) and~(\ref{zacc}) for the four parameters $\alpha b^2$,
$\rho_{s0}$, $\alpha d^2$ and $\Lambda$ of our model.

Note that from Eqs.~(\ref{qq}) and~(\ref{zacc}) we have that
\begin{eqnarray}\label{oD}
\rho_{D0}=\al(b^2-d^2)=K\rho_{s0},
\end{eqnarray}
where
\begin{eqnarray}\label{K}
K=\frac{4(1+q_0)\bar z^2-2(2-q_0)}{(1-2q_0)-(1+q_0)\bar z}, \quad
\bar z=(1+z_{acc})^3.
\end{eqnarray}
Since $\rho_{D0}$ is related to the energy density of the CDF, we
must require that $K>0$. So from Eqs.~(\ref{om}), (\ref{ox}),
(\ref{qq}) and~(\ref{zacc}) we have that
\begin{eqnarray}\label{alpha}
\al b^2=\n{al} 3H_0^2\om_{m,0},
\end{eqnarray}
\begin{eqnarray}
\rho_{s0}=\frac{6H_0^2}{K(2+\bar z)+2(1+2\bar z^2)},
\end{eqnarray}
\begin{eqnarray}\label{d}
d^2= b^2-\frac{6 K H_0^2}{\alpha [K(2+\bar z)+2(1+2\bar z^2)]},
\end{eqnarray}
\begin{eqnarray}\label{Lambda}
\Lambda=\frac{3H_0^2\bar z(K+4\bar z)}{K(2+\bar z)+2(1+2\bar z^2)},
\end{eqnarray}
where we have used Eqs.~(\ref{omega igual 1}) and~(\ref{oD}).

Now the four model parameters need to be constrained. We do this by
using the Eqs.~(\ref{alpha})--(\ref{Lambda}) and by considering the
increasing bulk of observational data that have been accumulated
during the past decade. The present expansion rate of the Universe
is measured by the Hubble constant. From the final results of the
Hubble Space Telescope Key Project~\citep{Freedman} to measure the
Hubble constant we know that its present value is constrained to be
$H_0 = 68 \pm 7$ km$s^{-1}$ Mpc$^{-1}$ ~\citep{Chen}. Equivalently
we can write $H_0^{-1}=9.776 \,  h^{-1}$ Gyr, where $h$ is a
dimensionless quantity and varies in the range $0.61 < h < 0.75$.

Now assuming a flat Universe, i.e. Eq.~(\ref{omega igual 1}) is
valid, Perlmutter et al.~\citep{Perlmutter} found that the
dimensionless density parameter $\Omega_{m,0}$ may be constrained to
be $\sim 0.3$,  implying from Eq.~(\ref{omega igual 1}) that
$\Omega_{x,0} \sim 0.7$~\citep{Copeland,Capozziello1,Capozziello2},
and the present day deceleration parameter $q_0$ may be constrained
to be $-1 < q_0 <
-0.64$~\citep{Copeland,Capozziello1,Capozziello2,Jackson}.

For consistency we also need to compare the age of the Universe
determined from our model with the age of the oldest stellar
populations, requiring that the Universe be older than these stellar
populations.  Specifically, the age of the Universe $t_0$ is
constrained to be $t_0 >$ 11-14
Gyr~\citep{Copeland,Jimenez1,Jimenez2,Jimenez3,Jimenez4,Jimenez5,Jimenez6,JimenezA}.

So let us calculate the age of the Universe from the Friedmann
equation~(\ref{frdeq}). This equation may be written as
\begin{eqnarray}\label{HH}
H^2=\frac{H_0^2}{a^6} \left( \frac{\rho_{s0}+ \alpha
a^3(b^2-d^2)+a^6\Lambda}{\rho_{s0}+\alpha(b^2- d^2)+\Lambda}
\right),
\end{eqnarray}
obtaining Eq.~(\ref{omega igual 1}) when the expression~(\ref{HH})
is evaluated today. Then the age of the Universe may be written as
\begin{eqnarray}\label{HHH}
&t_0=\int_0^{\infty} \frac{dz}{H(1+z)}=\nonumber &\\
&\frac{1}{H_0}\int_0^{\infty}
\sqrt{\frac{\alpha(b^2-d^2)+\rho_{s0}+\Lambda}{\alpha(b^2-d^2)(1+z)^3+
\rho_{s0}(1+z)^6+\Lambda}}\,\,\,\frac{dz}{(1+z)}. &
\end{eqnarray}

 In the Table~\ref{tabla15} we include some values obtained from
Eqs.~(\ref{alpha})--(\ref{Lambda}) for the parameters $\alpha b^2$,
$\alpha d^2$, $\rho_{s0}$, $\Lambda$ and $K$ corresponding to some
given values of the parameters $H_0$, $q_0$ and $z_{acc}$ (with
$c=1$ and $G=1$). For the Hubble parameter $H_0$ is considered both
possible values $H_{0-}$ for $h=0.61$ and $H_{0+}$ for $h=0.75$, and
for the matter dimensionless density parameter we have taken
$\Omega_{m,0}=0.3$. The last two columns represent the age of the
Universe determined from the model parameters for $h=0.75$ and
$h=0.61$ respectively. Clearly in the proposed model there are
configurations which are allowed, being their ages larger than the
oldest known stellar ages, since there exist combinations of the
parameters $\alpha b^2$, $\alpha d^2$, $\rho_{s0}$, $\Lambda$ which
give $t_0
>$ 11-14 Gyr satisfying the stellar population constraints.

\begin{table*}
\small \caption{In this table we show some values of the model
parameters obtained for given $H_0$, $q_0$ and $z_{acc}$
($\Omega_{m,0}=0.3, c=1, G=1$). \label{tabla15}}
\begin{tabular}{@{}crrrrrrrrr@{}}
\tableline $H_0 \, (s^{-1})$ & $q_0$ & $Z_{acc}$ & $\alpha b^2$
($s^{-2}$) &$\alpha d^2 \, (s^{-2})$ & $\rho_{s0} \, (s^{-2})$  &
$\Lambda \, (s^{-2})$ & K & $H_{0+}, t_0$& $H_{0-}, t_0$\\
\tableline
$H_{0+}$ & -0.68 & 0.587 & $5.315 \times 10^{-36}$ & $2.009\times 10^{-36}$ & $ 2.368 \times 10^{-37}$ & $2.655 \times 10^{-35}$ & 14 & 12 (Gyr)& 15 (Gyr)\\
$H_{0-}$ & -0.68 & 0.587 & $3.516 \times 10^{-36}$ & $1.329 \times 10^{-36}$ & $1.566 \times 10^{-37}$ & $1.756 \times 10^{-35}$ & 14 & 12 (Gyr)& 15 (Gyr)\\
$H_{0+}$ & -0.68 & 0.94 & $5.315 \times 10^{-36}$ & $1.538 \times 10^{-36}$& $1.415 \times 10^{-39}$ & $1.210 \times10^{-34}$ & 2619 & 11 (Gyr)& 13 (Gyr)\\
$H_{0-}$ & -0.68 & 0.94 & $3.516 \times 10^{-36}$ & $ 1.017 \times 10^{-36}$ & $9.361 \times 10^{-40}$ & $8.006 \times 10^{-35}$ & 2619 & 21 (Gyr)& 25 (Gyr) \\
$H_{0+}$ & -0.9 & 0.6 & $5.315 \times 10^{-36}$ & $5.126 \times 10^{-36}$ & $4.961\times 10^{-37}$ & $1.751 \times 10^{-35}$ & 0.38 & 11 (Gyr)& 13 (Gyr)\\
$H_{0-}$ & -0.9 & 0.6 & $3.516 \times 10^{-36}$ & $3.391 \times 10^{-36}$ & $3.281 \times 10^{-37}$ & $1.158 \times 10^{-35}$ & 0.38 & 11 (Gyr)& 13 (Gyr) \\
$H_{0+}$ & -0.9 & 1 & $5.315 \times 10^{-36}$  & $4.333 \times 10^{-36}$ & $9.926 \times 10^{-38}$ & $2.217 \times 10^{-35}$ & 9.9 & 14 (Gyr)& 17 (Gyr)\\
$H_{0-}$ & -0.9 & 1 & $3.516 \times 10^{-36}$ & $2.866 \times 10^{-36}$ & $6.566 \times 10^{-38}$ & $1.466 \times 10^{-35}$ & 9.9 & 14 (Gyr) & 17 (Gyr)\\
\tableline
\end{tabular}
%%%%% Any table notes must follow the \end{tabular} command.
%\tablenotetext{a}{Sample footnote for table~\ref{tbl-2} that was
%generated with the \LaTeX\ table environment} \tablenotetext{b}{Yet
%another sample footnote for table~\ref{tbl-2}}
%\tablenotetext{c}{Another sample footnote for table~\ref{tbl-2}}
%\tablecomments{We can also attach a long-ish paragraph of
%explanatory material to a table}
\end{table*}

%Note that if $d=0$ the age of the universe determined from our
%model is always less than those values given in the table for $d
%neq 0$.

%%%%%%%%%%%%%%%%%%%%%%%%%%%%%%%%%%%%%%%%%%%%%%%%%%%%%%
\subsection{The effect of the d--parameter}
%%%%%%%%%%%%%%%%%%%%%%%%%%%%%%%%%%%%%%%%%%%%%%%%%%%%%%
Now it is interesting to get some insights concerning the nature of
the $d$--parameter for studying the effect on the cosmological
evolution of the negative term $-\alpha d^2/a^3$ in the
Eq.~(\ref{rho}) for the dark energy component $\rho_x$. It can be
shown that in the proposed cosmological scenario the dominant energy
condition (DEC) is violated thanks to the presence of this
parameter. Effectively, if $d=0$ the dark energy state parameter
$w_x$ and the state parameter of the full source content $w_{_T}$
are given by
\begin{eqnarray}
w_x=\frac{\rho_{s0}-\Lambda a^6}{\rho_{s0}+\Lambda a^6}, \,\,\,\,\,
w_{_{T}}=\frac{\rho_{s0}-\Lambda a^6}{\rho_{s0}+\Lambda a^6+\alpha
b^2 a^3},
\end{eqnarray}
respectively. From these expressions we see that always $-1<w_x<1$
which implies that now the dark energy component satisfies DEC, as
well as $w_{_{T}}$. Their general behavior is shown in
Fig.~\ref{dcero} (compare with Fig.~\ref{stateP}).
\begin{figure}
\plotone{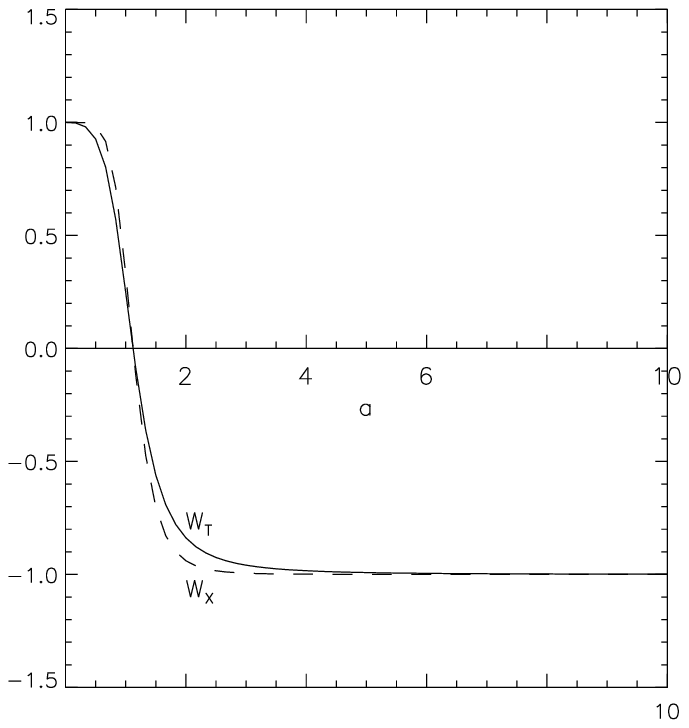}% Here is how to import EPS art
\caption{\label{dcero} We show the behavior of $w_x$ and of $w_{_T}$
as a functions of the scale factor. The dashed line represents the
behavior of the dark energy state parameter, while the solid line
represents the behavior of the state parameter of the full source
content. Both satisfy the DEC since in this case $d=0$.}
\end{figure}

However, the fulfilment of the DEC does not imply that the Universe
has a decelerated expansion. From Eq.~(\ref{tresp}) we may write
that
\begin{eqnarray}\label{r3p}
\frac{6\ddot{a}}{a}=  -(\rho_{_T}+3p_{_T})= \nonumber \\- \left(
\frac{4\rho_{s0}}{a^6}+\frac{\alpha (b^2-d^2)}{a^3}-2\Lambda
\right),
\end{eqnarray}
and putting $d=0$ we see that the accelerated expansion is realized
if $a > a_{acc}=((\alpha b^2 +\frac{\sqrt{\alpha^2 b^4+32 \Lambda
\rho_{s0}})}{4\Lambda})^{1/3}$. Another property of the d--parameter
to be considered is its effect on the deceleration parameter $q_0$.
From Eq.~(\ref{d}), which is independent of the Hubble parameter
$H_0$ and, using Eqs.~(\ref{K}) we can express the deceleration
parameter $q_0$ as a function of the $z_{acc}$ obtaining
\begin{eqnarray}
q_0(z_{acc})=\frac{(\bar z-1)\left( (2
{\bar z}+1)(b^2-d^2)\alpha -4H_0^2 (\bar z+1) \right)}{2 H_0^2(2\,{\bar z}^{2}+1)}, \nonumber \\
\end{eqnarray}
where, as before, $\bar z=(1+z_{acc})^3$. We can see that in this
case $q_0(z_{acc})$ rapidly tends to the value
\begin{eqnarray}
q_{0,\infty} = -1+\frac{\alpha}{2H_0^2}(b^2-d^2)=-1+\frac{3
\Omega_{m,0}}{2}\left( 1-\frac{d^2}{b^2}\right), \nonumber \\
\end{eqnarray}
obtaining, for $d=0$ and $\Omega_{m,0}=0.3$, the value
$q_{0,\infty}=-0.55$ and, from this value the deceleration parameter
reaches the value $q_{0,\infty} =-1$ for $d \approx b$(see
Fig.~\ref{fig15zacc}).
\begin{figure}
\plotone{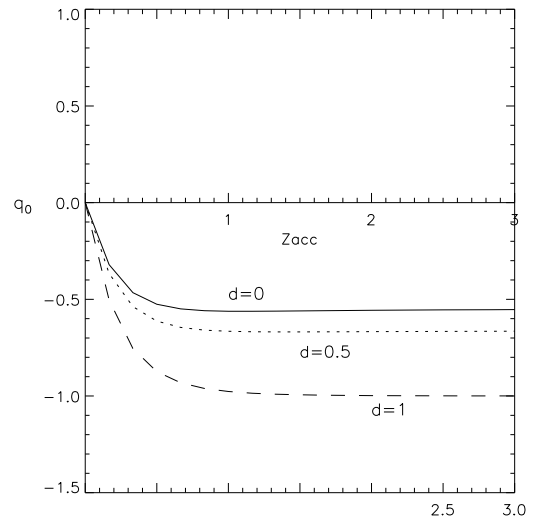}% Here is how to import EPS art
\caption{\label{fig15zacc} We show the behavior of the deceleration
parameter $q_0$ as a function of $z_{acc}$ for three values of the
parameter $d$ $(0,1/2,1)$ with $\Omega_{m,0}=0.3$ and $b=1$. We can
see that the deceleration parameter rapidly tends to the values
$(-0.55,-0.6625,-1)$ respectively.}
\end{figure}
So the parameter $d$ directly affects the range of validity of the
deceleration parameter which is constrained to be in the range $-1 <
q_0 <0$. Note that the value $q_{0,\infty}=-1$ for $d=b$ is
independent of the value of the dimensionless density parameter
$\Omega_{m,0}$.

%\newpage

%%%%%%%%%%%%%%%%%%%%%%%%%%%%%%%%%%%%%%%%%%%%%%%%%
\section{Concluding remarks}
%%%%%%%%%%%%%%%%%%%%%%%%%%%%%%%%%%%%%%%%%%%%%%%%%
Since today the observations constrain the value of $\omega$ to be
close to $\omega=-1$, we have considered broader cosmological
scenarios in which the equation of state of dark energy changes with
time. The two principal ingredients of the model are a stiff fluid
which dominates at early time~\citep{Dutta} and a CDF. The positive
part of the latter was associated with a dark matter component while
its negative part was considered as a part of the dark energy
component and was responsible for the effective dark energy density
crossing the phantom divide (see and compare Figs.~\ref{stateP}
and~\ref{dcero}). At the end, this cosmological model becomes
accelerated recovering the standard $\Lambda$CDM cosmology.

In general, the model may be seen as a continuation of the inflation
era. In the inflationary paradigm the scalar field $\Phi$, driven by
the potential $V(\Phi)$, generates the inflationary stage. In the
slow roll limit $\dot{\Phi}^2<< V(\Phi)$, with
$\omega_{\Phi}\approx-1$, we have a superluminal expansion while in
the kinetic--energy dominated limit $\dot{\Phi}^2>> V(\Phi)$, with
$\omega_{\Phi}\approx 1$, we have a stiff matter scenario
characterized by a subluminal expansion. Taking into account that
our model has a variable equation of state, we can think of it as a
transient model which interpolates smoothly between different
barotropic eras as, for instance, radiation dominated era, matter
dominated era an so on. In other words,  from
Eqs.~(\ref{omegax})--(\ref{r}) we see that $w_x \rightarrow 1$ (and
$w_{_T}\rightarrow 1$) for $a \rightarrow 0$ implying that the
energy density $\rho_x$ behaves like $1/a^6$ and matching, after
inflation, with the kinetic--energy mode of the scalar field
$\rho_{\Phi}\propto 1/a^6$. Now from Eq.~(\ref{omegax}) we see that
$\rho_x$ passes through a radiation dominated stage (i.e.
$\omega_x=1/3$) for
%\begin{eqnarray}
%a= \left(\frac{\alpha\,{d}^{2}+\sqrt
%{{\alpha}^{2}{d}^{4}+32\,\Lambda{ \rho_{s0}}}}{8\Lambda}\right)^{1/3}>0,
%\end{eqnarray}
\be
a_{rad}=\frac{a_m}{2^{2/3}}\left[\left(\frac{a_m}{a_c}\right)^{3}+
\sqrt{8+\left(\frac{a_m}{a_c}\right)^{6}}\,\right], \ee behaving
like $\rho_x \approx 1/a^4_{rad}$ and dominating over $\rho_m$.
After that, at $a=a_m$, we have $w_x=0$ and the effective dark
component behaves as a pressureless source thus obtaining a
matter--dominated stage. Finally the model evolves from this state
to a vacuum-energy dominated scenario. It is interesting to note
that in the a matter--dominated stage, if the
condition~(\ref{constraint}) is fulfilled, then the total energy
density is given by $\rho_{_T}=\alpha(b^2-d^2)/a^3_m+2\Lambda$ and,
if $a_m=a_c$ then $\rho_{_T}=\alpha b^2/a^3_m$, being $\rho_x=0$,
implying that we have at this stage only the dark matter component.

The above results indicate that a cosmological scenario based on a
CDF component and the effective multifluid configuration $\rho_x$
can, in certain cases, reproduce the quintessential behavior (see
Figs.~\ref{stateP} and~\ref{dcero}). In fact, the state parameter of
the total matter content $-1<w_{_T}<1$ is constrained the same as is
the state parameter of the scalar field  in quintessence models. In
this manner we avoid the use of scalar fields and particular classes
of potentials for describing the dark energy component.

Finally, all the parameters of the model have been expressed in
terms of the observable quantities which may be constrained by the
astrophysical observational data. In effect, in Table~\ref{tabla15}
some values of the model parameters $\alpha b^2$, $\alpha d^2$,
$\rho_{s0}$ and $\Lambda$ were included, which correspond to some
given values of the parameters $H_0$, $q_0$, $z_{acc}$ and
$\Omega_{m,0}$ constrained by astrophysical observations.

\section{acknowledgements}
LPC acknowledges the hospitality of the Physics Department of
Universidad del B\'\i o-B\'\i o where a part of this work was done.
The authors thank Paul Minning for carefully reading this
manuscript. This work was partially supported by CONICYT through
grants FONDECYT N$^0$ 1080530 (MC), MECESUP USA0108 (MC and LPC) and
Project X224 by the University of Buenos Aires along with Project
5169 by the CONICET (LPC). It was also supported by the Direcci\'on
de Investigaci\'on de la Universidad del B\'\i o--B\'\i o (MC) and
Consejo Nacional de Investigaciones Cient\'\i ficas y T\'ecnicas
(LPC).

%\nocite{*}
%\bibliographystyle{spr-mp-nameyear-cnd}
%\bibliography{myref}

\end{document}